\documentclass[twocolumn,amsmath,amssymb,floatfix,prl]{revtex4-1}

\usepackage{graphicx}
\usepackage{bm}
\usepackage{subfigure}
\usepackage{natbib}
\usepackage{layout}
\usepackage{setspace}
\usepackage{psfrag}
\usepackage{epstopdf}
\usepackage[colorlinks,citecolor=blue,linkcolor=blue]{hyperref}
\usepackage[all]{hypcap}
\usepackage{xcolor}
\usepackage{appendix}

\newcommand{\Ca}{\mathrm{Ca}}

\newcommand{\dhat}{$\hat{d}\;$}
\newcommand{\nhat}{$\hat{n}\; $}
\newcommand{\dhata}{$\hat{d}$}

\newcommand{\gdota}{\dot{\gamma}}

\begin{document}

\title{Mechanism of margination in confined flows of blood and other \\ multicomponent suspensions}

\author{Amit Kumar}
\author{Michael D. Graham}%
\email{graham@engr.wisc.edu}
\affiliation{Department of Chemical and Biological Engineering, University of Wisconsin-Madison, Madison, Wisconsin 53706, USA}
\date{\today}
\begin{abstract}
\noindent Flowing blood displays a phenomenon called margination, in which leukocytes and platelets 
are preferentially found near blood vessel walls, while erythrocytes are depleted from these regions.
Here margination is investigated using direct hydrodynamic simulations of a binary suspension of stiff (s) and floppy (f) capsules, as well as a stochastic model that incorporates the key particle transport mechanisms in suspensions -- wall-induced hydrodynamic migration
and shear-induced pair collisions. The stochastic model allows the relative importance of these two mechanisms to be directly evaluated and thereby indicates that margination, at least in the dilute case, is largely due to the differential dynamics of homogeneous (e.g. s-s) and heterogeneous (s-f) collisions. 
\end{abstract}
\maketitle

\textit{Introduction}---
Blood is a multicomponent suspension consisting primarily of red blood cells (RBCs)
along with trace amounts of other components, primarily leukocytes and platelets \citep{popel05,kumar_rev}.
Under physiological flow conditions, both the leukocytes and platelets segregate 
near the vessel walls \citep{tangelder85,*lipowsky89}, a phenomenon known as margination, while the RBCs
tend to be depleted in the near-wall region forming a so-called ``cell-free layer'' \citep{popel05,kumar_rev}. 
Leukocytes are larger than RBCs and platelets smaller, but  both are considerably stiffer than
RBCs \citep{freund07,*zhao11,*fedosov12}; this difference is believed to play an important role in their margination.
In addition, in illnesses such as malaria and sickle cell disease, the RBCs themselves
are known to become stiff, and these stiffened cells also marginate \citep{hou10}. Furthermore, delivery of drugs to tumors using particles injected into the bloodstream is a major 
goal of current cancer research \citep{loomis10,*cabral11}; margination may influence the distribution and thus the efficacy of these particles. Finally, there are many microfluidic
applications where differences in the margination properties of various 
blood components are exploited to effect their separation \citep{munn05,hou10,kumar_rev}.

Despite the importance of flow induced segregation and margination phenomena
in particle mixtures like blood, a mechanistic understanding is elusive \cite{kumar_rev}. 
The goals of the present work are to establish the mechanisms of rigidity-based
margination in confined flows and to illustrate that these mechanisms are generic for multicomponent 
suspensions. We do this in two parts. 
First, we consider direct hydrodynamic simulations of a model problem 
that isolates the effect of stiffness on margination: a binary suspension of fluid-filled elastic
capsules subjected to simple shear flow in a planar slit \citep{kumar_seg}. The two components 
of the binary mixtures have unequal membrane rigidities -- the component with the higher rigidity
is termed stiff, while the component with the lower rigidity is termed floppy. 
Additionally, we employ an idealized master equation (ME) model of the suspension
dynamics that incorporates the two key transport mechanisms in
confined suspensions: (1) wall-induced particle migration and (2) particle displacements
in homogeneous (e.g. stiff-stiff) and heterogeneous (stiff-floppy) pair collisions. We introduce a novel hydrodynamic Monte Carlo (HMC) simulation technique to find
steady state concentration distributions for this model. In contrast to direct numerical 
simulations, the ME/HMC approach allows the various aspects of the particle dynamics to
be independently controlled, thereby enabling delineation of their role in the segregation behavior. 

Using these approaches, we are able to isolate the effect of heterogeneous pair collisions and thus demonstrate their dominant role, at least in dilute systems, in the observed margination behavior.  
The approach and mechanisms presented here are generic for multicomponent suspensions -- the model only takes as inputs the migration behavior of the various components of the suspension and their displacements upon homogeneous and heterogeneous collisions. Therefore, it is extensible to other systems including mixtures of particles of unequal sizes and shapes, thus encompassing whole blood flow as well as flow of drug delivery particles in the bloodstream.

\begin{figure}[!t]
\centering
\includegraphics[width=0.4\textwidth]{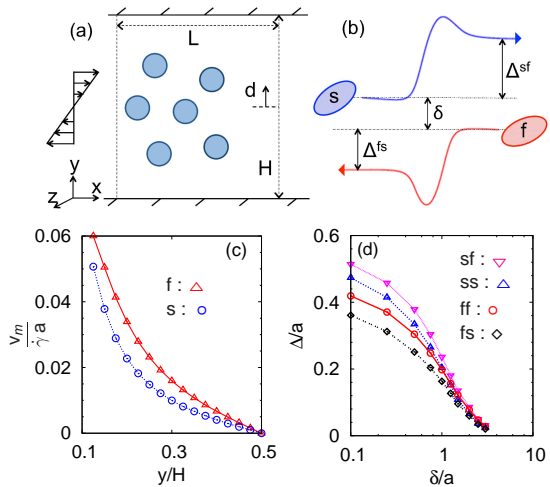}
\caption{(color online) (a) System geometry, (b) schematic of particle trajectories in a pair collision, 
(c) isolated particle migration velocity at confinement ratio $2a/H=0.197$, and (d) cross-stream displacement $\Delta$ in various types of pair collisions as a function of the
initial offset $\delta$. Here `s' refers to the stiffer particle ($\Ca_s=0.2$), while `f' refers to the floppier particle ($\Ca_f=0.5$).
}\label{fig:geom} 
\end{figure}

\textit{Formulation} ---
We consider a suspension of fluid-filled neo-Hookean capsules \citep{kumar_seg} subjected to simple shear flow  with shear rate $\dot{\gamma}$
between infinite parallel walls at $y=0$ and $y=H$ (Fig. \ref{fig:geom}a). 
At rest, all capsules are spheres with the same radius $a$. 
The rigidity of a particle is characterized by its membrane shear modulus $G$, which is expressed 
in terms of the non-dimensional capillary number $\Ca=\mu\dot{\gamma}a/G$, where $\mu$ is the 
viscosity of the suspending fluid. The capsules in a binary mixture with a lower $\Ca=\Ca_s$ are termed stiff, while the capsules with a higher $\Ca=\Ca_f$ are termed floppy; the number fraction of the floppy particles will be denoted $X_f$. The viscosity ratio $\lambda$ of the fluid inside and outside the particle is unity. The Reynolds number 
is taken to be negligible.

In a dilute suspension of particles, particle interactions can be treated 
as a sequence of uncorrelated pair collisions \citep{zurita12,cunha96}.
In addition, since the capsules are deformable, they also have a wall-induced migration
velocity $v_m(y)$  away from the wall \citep{leighton91,*graham05}. The effect of the
 pair collisions and the wall-induced particle migration can be consistently
described by a kinetic master equation (cf.~\cite{zurita12}). For simplicity, we focus here on a model of a monolayer in the $x-y$ (flow-gradient) plane  -- extension to three dimensions is straightforward  \citep{zurita12}. 
The mean area number density of all the particles in the monolayer is denoted $n_0$, such that the areal fraction in the $x-y$ plane at rest is $\phi_a = \pi a^2 n_0$. The mean area number density of each of the species $\alpha$ in the mixture is denoted $n_0^\alpha$, while its distribution in the $y$ direction is denoted $n^{\alpha}(y)$. In this case, the master equation is:
\begin{multline}\label{eq:meq}
\frac{\partial n^{\alpha}(y)}{\partial t} = - \frac{\partial \big(v_m^{\alpha} n^{\alpha}\big)}{\partial y}  +  \sum_{\beta=1}^{N_s} \bigg( \int_{y-H}^{y}  \big[ n^{\alpha}(y-\Delta^{\alpha\beta}) \\
\times n^\beta(y-\Delta^{\alpha\beta}-\delta) ~ -  ~n^\alpha(y)~n^\beta(y-\delta) \big] \, \dot{\gamma} \, |\delta| \, d\delta \bigg),
\end{multline}
where $\delta$ is the pre-collision pair offset in the $y$ direction, $\Delta^{\alpha\beta}(\delta)$ is
the cross-stream displacement of particle of type $\alpha$ after collision with another
particle of type $\beta$, while the sum is over all the
species $N_s$ in the suspension; see (Fig.~\ref{fig:geom}b) for a schematic of a pair collision. 
The first term on the right hand side arises from the 
wall-induced migration, while  the integral term represents the effect of pair
collisions \citep{zurita12}.

This equation is analogous to the Boltzmann equation for rarefied gases \citep{bird94,ivanov98}. The dynamic simulation Monte Carlo (DSMC) approach is a popular technique to obtain solutions of the Boltzmann equation \citep{bird94,ivanov98}. The current work is inspired by the DSMC approach, which has also found interest in recent works on colloidal suspensions of rigid spheres \citep{zurita12}. By analogy, we term the  method in the present work as the hydrodynamic Monte Carlo (HMC) method. As in the case of DSMC \citep{bird78,bird94}, the HMC approach is appropriate
in the dilute limit and requires the assumption of chaotic particle dynamics, an assumption that
is valid for the particulate flows considered here \citep{Drazer:2002gn}.

In the HMC approach, the $y$-positions of $N_p$  particles are followed in time. 
We set $N_p=100$ here; simulations with larger $N_p$ gave indistinguishable results.
Each particle is assumed to represent an infinite number of particles 
at the same $y$ position randomly distributed in the flow direction with an average spacing of $L$,
where $L$ is given by $L=N_p/(n_0 H)$. A distinguishing feature of the HMC (or DSMC) approach
is that the collisions between particles are treated probabilistically \citep{bird78}.
In the present study, we neglect pair collisions with large initial offsets $\delta>\delta_{\textrm{cut}}$, as their effect on cross-stream displacement is weak; we take $\delta_{\textnormal{cut}} = 2.5a$ here.
A timestep of the simulation involves choosing a pair of particles which satisfies the condition
$\delta \leq \delta_{\textnormal{cut}}$. The pair is subsequently selected or rejected for collision with a probability proportional to their relative velocity of approach $\gdota |\delta|$ \citep{bird94,koura86,zurita12}.
An important aspect of the simulation is the time interval between collisions $\Delta t$, because the
wall-induced particle migration occurs simultaneously with the collisions. In order to determine
$\Delta t$ at each time step, we assume that the number of particle collisions with time follows
a Poisson process with a mean collision frequency $\nu$ ($\nu$ is estimated \footnote{$\nu = 0.5 \int_{0}^{L} \int_{0}^{H} n(y) (\int_{y-\delta_{\textnormal{cut}}}^{y+\delta_{\textnormal{cut}}} n(y+\delta) \, \dot{\gamma} |\delta| \, d \delta ) \, dy \, dx$}),
such that the time interval between collisions is distributed with probability
$P(\Delta t) = \nu e^{-\nu \Delta t}$ \citep{koura86}.
The distribution of time interval between collisions results from the variation of particle positions in the flow direction, which is not explicitly specified. Once the collision pair and time interval are set, we update positions of all particles $k$ as: $y_k(t+\Delta t) = y_k(t) ~+~  v_m \Delta t ~+~ \Delta_k$, where $\Delta_k$ is non-zero only for
the colliding particle pair $(i,j)$. If the collision is homogeneous, then $\Delta_{k}=\Delta^{\textit{ss}}$ or $\Delta^{\textit{ff}}$ for stiff and floppy particles, respectively. If the collision is heterogeneous, then $\Delta_{k}=\Delta^{\textit{sf}}$ for the stiff particle and $\Delta_{k}=\Delta^{\textit{fs}}$ for the floppy one. The procedure outlined
above is repeated until a statistical steady state is obtained.
Other details of the method can be found in \citet{koura86}, whose approach is closely followed here.

\textit{Pair collisions and wall induced migration.}---
The HMC method requires as inputs the cross-stream displacements $\Delta^{\alpha\beta}$ in pair collisions and the wall-induced migration velocity $v_{m}^{\alpha}$. These were computed using an accelerated boundary integral 
method \citep{kumar_seg,kumar_jcp}; also see \citep{supmat}.
Figure (\ref{fig:geom}c) shows the isolated particle migration velocity as a function of $y/H$
for $\Ca_s=0.2$ and $\Ca_f=0.5$ capsules at a confinement ratio $2a/H=0.197$.   
Results for $\Delta^{\alpha\beta}(\delta)$ for the same two species (in an unbounded domain) are shown
in Fig. (\ref{fig:geom}d). For offsets $\delta\lesssim a$, $\Delta^{\textit{ff}} < \Delta^{\textit{ss}}$. Furthermore, an important feature in this plot, first reported in Ref.~\cite{kumar_seg}, is that the displacement of the stiffer particle in heterogeneous collisions is higher than that of the floppy particle ($\Delta^{\textit{sf}}>\Delta^{\textit{fs}}$), while the cross-stream displacements in homogeneous collisions are between these two limits, i.e., $\Delta^{\textit{fs}} < \Delta^{\textit{ff}} , \Delta^{\textit{ss}} <\Delta^{\textit{sf}}$. 
This ordering will turn out to be crucial in determining the segregation behavior. 

The above particle migration velocity and pair collision results were determined in idealized 
systems, namely, by considering an isolated particle and an unconfined system, respectively.
Our interest is in a \textit{confined suspension}, so we expect that corrections will be necessary to both the migration and the pair collision results. For the migration velocity $v_m$, 
we note the wall-induced migration is due to the disturbance velocity created by the particle's image, and is a far-field effect \citep{leighton91,*graham05}. As a result of 
its far-field nature, in a suspension of particles, it can be expected that the wall-induced 
migration of a particle will result not only due to interaction with its own images, but also due
to images of other particles -- this can be expected to introduce an averaging effect in a suspension of particle mixtures.
To model this, we modify the migration velocities as per the following equation: 
$v_m^{\alpha,s} = \xi \, v_m^{\alpha} ~ +  (1-\xi) \, \sum_{1}^{N_s} X_{\beta} v_m^{\beta}$,
where $v_m^{\alpha,s}$ is the migration velocity of the species $\alpha$ in the suspension and $\xi$ is an adjustable parameter. By taking $\xi=1$, we recover the isolated particle migration velocity for 
each species, while for $\xi=0$, each of the species in the suspension has
the same migration velocity.

We adopt a similarly simple model to account for the confinement effects on cross-stream
displacement $\Delta$ in pair collisions. For a spherical particle with center at $y=a$, 
its $\Delta$ in a pair collision is expected to be zero as it will be touching the wall,
while its $\Delta$ will approach the unconfined result at large particle-wall separations.
To account for this effect, we multiply the $\Delta$ of the particle in an
unconfined system by a factor $\eta = 1 - e^{-(d_w - a)/a}$, where $d_w$
is the distance of the particle from the nearest wall, assumed to satisfy $d_w > a$.
Despite this correction factor, particles might still occasionally overlap with the walls
at high volume fractions, though that did not occur in the regime investigated here. 

\begin{figure}[!t]
\centering
\includegraphics[width=0.48\textwidth]{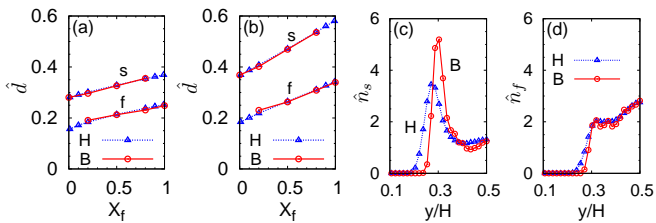}
\caption{(color online) (a) and (b): Mean normalized distance of a species $\hat{d}=2\bar{d}/H$ from the centerline in HMC (H) and BI (B) methods in (a) $(\Ca_s,\Ca_f,\phi_v)=(0.2,0.5,0.04)$ mixture as a function of $X_f$ and in (b) $(0.2,0.5,0.12)$ mixture.  
(c) and (d): Normalized number density profile \nhat for (c) stiff and (d) floppy particles at $X_f=0.5$ for the suspension in (a).
In all cases $2a/H=0.197$.}\label{fig:yavg1}
\end{figure}

\textit{Validation.}---We next validate the HMC method by comparing it with the results of detailed boundary integral (BI) simulations \citep{kumar_seg,kumar_jcp}.
The suspensions in the BI simulations are fully three dimensional, but otherwise
are similar to those in the HMC method. The particle volume fraction in the
BI simulations will be denoted $\phi_v$.
The HMC model has two adjustable parameters, $n_0$ (or $\phi_a$) and $\xi$,
which can be tuned to obtain good agreement with BI simulations for a
given suspension. In the present work, we seek the agreement
of the mean normalized distance of a species from the centerline 
$\hat{d} = 2\bar{d}/H$ (see Fig. \ref{fig:geom}a) between the two methods.

The parameters $n_0$ and $\xi$ in the HMC method are expected to depend on $\phi_v$ and $2a/H$,
while their dependence on $\Ca_s$, $\Ca_f$ and $X_f$ is expected to be weak. To demonstrate this,
we focus on suspensions with $\phi_v=0.04$ and $2a/H=0.197$ held fixed. We then consider the BI results for a pure suspension with $\Ca=0.2$, and tune the value of $n_0$ in the HMC method
to obtain a good match in \dhat between the two methods; this yields $n_0=0.026a^{-2}$ (or $\phi_a=0.082$). We next consider BI results for a binary mixture with $\Ca_s=0.2$, $\Ca_f=0.5$, and $X_f=0.5$ and tune the value of $\xi$ in the HMC method to obtain a good agreement in \dhat of both the species between the two methods;  this yields $\xi=0.23$. Having set the value of $n_0$ and $\xi$, the HMC method can then be used to \textit{predict} results for other suspensions at the same $\phi_v$ and $2a/H$. To establish this, we consider the same binary mixture as above ($\Ca_s=0.2$, $\Ca_f=0.5$) and predict \dhat for both the species for a range of $X_f$ and compare them with the corresponding BI results (Fig. \ref{fig:yavg1}a). Excellent agreement can be observed at all values of $X_f$. Similar close agreement was also observed for different sets of
($\Ca_{s},\Ca_f$), namely  (0.1, 0.5) and (0.3, 0.4) with no adjustment of $n_0$ and $\xi$ (see \citep{supmat}). Lastly, we consider a suspension with $(\Ca_s, \Ca_f)$ of (0.2, 0.5) at $\phi_v=0.12$ and $2a/H=0.197$, and we tune the values of $n_0$ and $\xi$ as discussed above, which yield $n_0=0.093 a^{-2}$ (or $\phi_a=0.292$) and $\xi=0.63$.
Subsequently, we predict \dhat at various $X_f$ in this system (Fig. \ref{fig:yavg1}b).
Excellent agreement is observed even at this higher volume fraction. 

Besides the averaged measure \dhata, we also compared the particle number density distribution in the wall normal direction in the HMC and BI simulations.
Results for the normalized number density distribution $\hat{n}_{\alpha}(y) = n_{\alpha}(y)/n_0^\alpha$ are shown in Figs. (\ref{fig:yavg1}c) and (\ref{fig:yavg1}d) for the $\Ca_s=0.2$ and $\Ca_f=0.5$ mixture at $\phi_v=0.04$, $2a/H=0.197$,
and $X_f=0.5$; a more complete set of plots is provided in \citep{supmat}. Very good agreement of \nhat  with the BI results is observed in all cases. The agreement is remarkably good in the region around the centerline, though the peak near the wall is usually smeared in the HMC results in comparison to the BI results. 
Nonetheless, given the broad agreement of \dhat as well \nhat in HMC and BI simulations in various suspensions, it is apparent that the HMC model captures the key aspects of the particle distributions in these suspensions.

\begin{figure}[!t]
\centering
\includegraphics[width=0.42\textwidth]{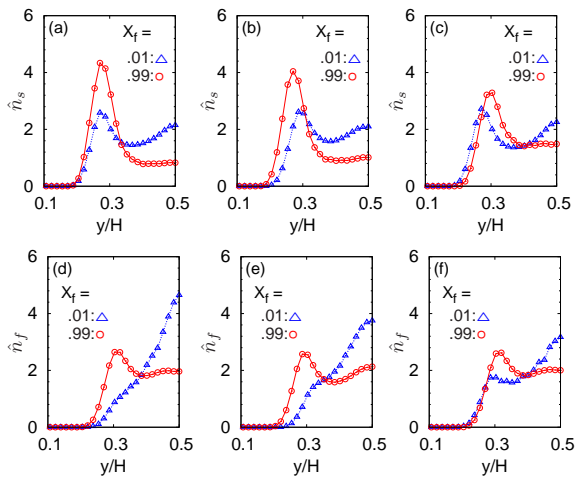}
\caption{(color online) Number density \nhat for stiff (top row) and floppy (bottom row)
particles in simulations at $X_{f}=0.01$ and $X_{f}=0.99$, from: (a) and (d), full model (case A);  (b) and (e), only difference between the 
two species being heterogeneous collisions (case B);   (c) and (f),  only difference between the two species being migration velocities (case C). These plots are for $(\Ca_s,\Ca_f,\phi_v)=(0.2,0.5,0.04)$.}\label{fig:nden}
\end{figure}

\textit{Mechanisms of flow induced segregation.}--- 
A key benefit of the HMC approach is that it allows independent 
investigation of the effect of various ingredients of the particle dynamics on the margination behavior. We focus here on suspensions with $\Ca_{s}=0.2$, $\Ca_{f}=0.5$, $\phi_v=0.04$, and $2a/H=0.197$. As discussed above, the HMC results for this system were generated by setting $n_0=0.026a^{-2}$ and $\xi=0.23$. We first show the plots for \nhat from the full HMC model
for both the stiff and floppy particles at  $X_f=0.01$ (dilute in floppy) and $X_{f}=0.99$ (dilute in stiff) in Figs. (\ref{fig:nden}a)
and (\ref{fig:nden}d), respectively. It is clear from these plots that the \emph{stiff} particles 
accumulate in the particle layer formed nearest to the wall as they become dilute in the suspension, i.e., they marginate with increasing $X_f$. In contrast,  the \emph{floppy} particles do the opposite as they become dilute ($X_{f}$ decreases), accumulating near the centerline  and thus ``antimarginating''.
These trends agree with the detailed boundary integral results of Ref.~\cite{kumar_seg}; also see \cite{supmat}.

To disentangle the effects of wall induced migration and pair collisions, we now consider a number of control cases. First, we investigate the impact of heterogeneous collisions, by (i) setting the particle migration velocities of both the species to the simple average migration velocity of these two species and (ii) setting $\Delta^{\textit{ss}}$ and $\Delta^{\textit{ff}}$ to the average value for the two species. Therefore, the only difference between these two species is their behavior in heterogeneous collisions: $\Delta^{\textit{sf}}>\Delta^{\textit{fs}}$. Plots of $\hat{n}(y)$ for these simulations are shown in Figs. (\ref{fig:nden}b) and (\ref{fig:nden}e). 
The difference between $\Delta^{\textit{sf}}$ and $\Delta^{\textit{fs}}$ is sufficient to lead to a segregation between the two species. In fact, as will be quantified shortly, most of the segregation results from heterogenous collisions. 

We next examine the effect of differences in migration velocity on the segregation behavior by 
  setting the cross-stream displacement in all types of collisions for both the 
species to the simple average of the four curves on Fig. (\ref{fig:geom}d),
yielding $\Delta^{\textit{sf}}=\Delta^{\textit{ss}}=\Delta^{\textit{fs}}=\Delta^{\textit{ff}}$. Plots for $\hat{n}(y)$ in this case
are shown in Figs. (\ref{fig:nden}c) and (\ref{fig:nden}f). Here too some  segregation is observed, though the degree of segregation is considerably smaller than in the full model.	
\begin{figure}[!t]
\centering
\includegraphics[width=0.45\textwidth]{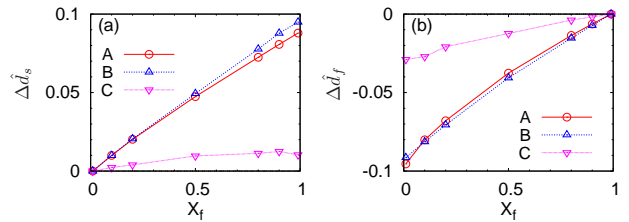}
\caption{(color online) The difference $\Delta \hat{d}$ between \dhat of a species in the mixture and the \dhat
	of that species in its pure suspension for (a) stiff particles
	and (b) floppy particles in $(\Ca_s,\Ca_f,\phi_v)$=$(0.2,0.5,0.04)$ suspension. The cases A-C are described in the text.}\label{fig:seg}
\end{figure}
	
The degree of segregation between the two species is more quantitatively characterized by computing
the difference in \dhat of each of the species from the corresponding pure species result; this is denoted by $\Delta \hat{d}$. The plots for $\Delta \hat{d}$ in various cases described above are shown in Figs (\ref{fig:seg}a) and (\ref{fig:seg}b) for the stiff and floppy particles, respectively.
For the present parameter set, the degree of segregation from the full model (case A) and the model where only heterogeneous collisions are distinct (case B)
are almost identical, while that resulting from differences in the migration velocity (case C) is much weaker. 
In a recent direct simulation study  \citep{kumar_seg}, heterogeneous collisions were conjectured to play an important role in margination. In the simulations, however, it is not possible to deconvolve the effects of migration and collisions: i.e.~one cannot perform control simulations. The ME/HMC approach does not suffer from this limitation.

Finally, we have investigated the effect of volume fraction on particle dynamics and margination. At $\phi_v=0.12$, the effects of heterogeneous collisions and differential migration velocities (cases B and C) on margination become comparable (see \citep{supmat}). Very similar results arise at $\phi_v=0.2$ \citep{supmat}. In the microcirculation, the  volume fraction of cells is between $\phi_v=0.1-0.25$ and it appears that in this regime, both the migration velocity and the heterogeneous collisions are playing an important role in the segregation between the various species.

\textit{Conclusions.}---
To gain a mechanistic understanding of margination in blood and other multicomponent suspensions, we have introduced an idealized description of the particle flow dynamics that incorporates the two key transport mechanisms
in confined suspensions: wall-induced migration and hydrodynamic pair collisions. The results for this model system clarify the important and previously underappreciated role played by heterogeneous collisions in the observed segregation behavior. Because differential behavior in heterogeneous collisions is generic for particles with contrasting shape or size as well as flexibility, the insights
presented here are also applicable
for other multicomponent suspensions. In particular, they have important implications in the design of drug
delivery particles for optimal vascular wall targeting or for separating trace components of blood in microfluidic devices.

\begin{acknowledgments}
This work is supported by NSF grants CBET-0852976 (funded under the American Recovery and Reinvestment Act of 2009) and CBET-1132579.
\end{acknowledgments}


\begin{thebibliography}{24}%
\makeatletter
\providecommand \@ifxundefined [1]{%
 \@ifx{#1\undefined}
}%
\providecommand \@ifnum [1]{%
 \ifnum #1\expandafter \@firstoftwo
 \else \expandafter \@secondoftwo
 \fi
}%
\providecommand \@ifx [1]{%
 \ifx #1\expandafter \@firstoftwo
 \else \expandafter \@secondoftwo
 \fi
}%
\providecommand \natexlab [1]{#1}%
\providecommand \enquote  [1]{``#1''}%
\providecommand \bibnamefont  [1]{#1}%
\providecommand \bibfnamefont [1]{#1}%
\providecommand \citenamefont [1]{#1}%
\providecommand \href@noop [0]{\@secondoftwo}%
\providecommand \href [0]{\begingroup \@sanitize@url \@href}%
\providecommand \@href[1]{\@@startlink{#1}\@@href}%
\providecommand \@@href[1]{\endgroup#1\@@endlink}%
\providecommand \@sanitize@url [0]{\catcode `\\12\catcode `\$12\catcode
  `\&12\catcode `\#12\catcode `\^12\catcode `\_12\catcode `\%12\relax}%
\providecommand \@@startlink[1]{}%
\providecommand \@@endlink[0]{}%
\providecommand \url  [0]{\begingroup\@sanitize@url \@url }%
\providecommand \@url [1]{\endgroup\@href {#1}{\urlprefix }}%
\providecommand \urlprefix  [0]{URL }%
\providecommand \Eprint [0]{\href }%
\providecommand \doibase [0]{http://dx.doi.org/}%
\providecommand \selectlanguage [0]{\@gobble}%
\providecommand \bibinfo  [0]{\@secondoftwo}%
\providecommand \bibfield  [0]{\@secondoftwo}%
\providecommand \translation [1]{[#1]}%
\providecommand \BibitemOpen [0]{}%
\providecommand \bibitemStop [0]{}%
\providecommand \bibitemNoStop [0]{.\EOS\space}%
\providecommand \EOS [0]{\spacefactor3000\relax}%
\providecommand \BibitemShut  [1]{\csname bibitem#1\endcsname}%
\let\auto@bib@innerbib\@empty
\bibitem [{\citenamefont {Popel}\ and\ \citenamefont
  {Johnson}(2005)}]{popel05}%
  \BibitemOpen
  \bibfield  {author} {\bibinfo {author} {\bibfnamefont {A.~S.}\ \bibnamefont
  {Popel}}\ and\ \bibinfo {author} {\bibfnamefont {P.~C.}\ \bibnamefont
  {Johnson}},\ }\href@noop {} {\bibfield  {journal} {\bibinfo  {journal} {Ann.
  Rev. Fluid Mech.}\ }\textbf {\bibinfo {volume} {37}},\ \bibinfo {pages} {43}
  (\bibinfo {year} {2005})}\BibitemShut {NoStop}%
\bibitem [{\citenamefont {Kumar}\ and\ \citenamefont
  {Graham}(2012{\natexlab{a}})}]{kumar_rev}%
  \BibitemOpen
   \bibfield  {author} {\bibinfo {author} {\bibfnamefont {A.}~\bibnamefont
  {Kumar}}\ and\ \bibinfo {author} {\bibfnamefont {M.~D.}\ \bibnamefont
  {Graham}},\ }\href@noop {} {\enquote {\bibinfo {title} {Margination and
  segregation in confined flows of blood and other multicomponent
  suspensions},}\ }\bibinfo {note} {Soft Matter (2012), doi:
  10.1039/C2SM25943E}\BibitemShut {NoStop}%
\bibitem [{\citenamefont {Tangelder}\ \emph {et~al.}(1985)\citenamefont
  {Tangelder}, \citenamefont {Teirlinck}, \citenamefont {Slaaf},\ and\
  \citenamefont {Reneman}}]{tangelder85}%
  \BibitemOpen
  \bibfield  {author} {\bibinfo {author} {\bibfnamefont {G.~J.}\ \bibnamefont
  {Tangelder}}, \bibinfo {author} {\bibfnamefont {H.~C.}\ \bibnamefont
  {Teirlinck}}, \bibinfo {author} {\bibfnamefont {D.~W.}\ \bibnamefont
  {Slaaf}}, \ and\ \bibinfo {author} {\bibfnamefont {R.~S.}\ \bibnamefont
  {Reneman}},\ }\href@noop {} {\bibfield  {journal} {\bibinfo  {journal} {Am.
  J. Physiol-Heart C}\ }\textbf {\bibinfo {volume} {248}},\ \bibinfo {pages}
  {H318} (\bibinfo {year} {1985})}\BibitemShut {NoStop}%
\bibitem [{\citenamefont {Firrell}\ and\ \citenamefont
  {Lipowsky}(1989)}]{lipowsky89}%
  \BibitemOpen
  \bibfield  {author} {\bibinfo {author} {\bibfnamefont {J.~C.}\ \bibnamefont
  {Firrell}}\ and\ \bibinfo {author} {\bibfnamefont {H.~H.}\ \bibnamefont
  {Lipowsky}},\ }\href@noop {} {\bibfield  {journal} {\bibinfo  {journal} {Am.
  J. Physiol-Heart C}\ }\textbf {\bibinfo {volume} {256}},\ \bibinfo {pages}
  {H1667} (\bibinfo {year} {1989})}\BibitemShut {NoStop}%
\bibitem [{\citenamefont {Freund}(2007)}]{freund07}%
  \BibitemOpen
  \bibfield  {author} {\bibinfo {author} {\bibfnamefont {J.~B.}\ \bibnamefont
  {Freund}},\ }\href@noop {} {\bibfield  {journal} {\bibinfo  {journal} {Phys.
  Fluids}\ }\textbf {\bibinfo {volume} {19}},\ \bibinfo {pages} {023301}
  (\bibinfo {year} {2007})}\BibitemShut {NoStop}%
\bibitem [{\citenamefont {Zhao}\ and\ \citenamefont {Shaqfeh}(2011)}]{zhao11}%
  \BibitemOpen
  \bibfield  {author} {\bibinfo {author} {\bibfnamefont {H.}~\bibnamefont
  {Zhao}}\ and\ \bibinfo {author} {\bibfnamefont {E.~S.~G.}\ \bibnamefont
  {Shaqfeh}},\ }\href@noop {} {\bibfield  {journal} {\bibinfo  {journal} {Phys.
  Rev. E.}\ }\textbf {\bibinfo {volume} {83}},\ \bibinfo {pages} {061924}
  (\bibinfo {year} {2011})}\BibitemShut {NoStop}%
\bibitem [{\citenamefont {Fedosov}\ \emph {et~al.}(2012)\citenamefont
  {Fedosov}, \citenamefont {Fornleitner},\ and\ \citenamefont
  {Gompper}}]{fedosov12}%
  \BibitemOpen
  \bibfield  {author} {\bibinfo {author} {\bibfnamefont {D.~A.}\ \bibnamefont
  {Fedosov}}, \bibinfo {author} {\bibfnamefont {J.}~\bibnamefont
  {Fornleitner}}, \ and\ \bibinfo {author} {\bibfnamefont {G.}~\bibnamefont
  {Gompper}},\ }\href@noop {} {\bibfield  {journal} {\bibinfo  {journal} {Phys.
  Rev. Lett.}\ }\textbf {\bibinfo {volume} {108}},\ \bibinfo {pages} {028104}
  (\bibinfo {year} {2012})}\BibitemShut {NoStop}%
\bibitem [{\citenamefont {Hou}\ \emph {et~al.}(2010)\citenamefont {Hou},
  \citenamefont {Bhagat}, \citenamefont {Chong}, \citenamefont {Mao},
  \citenamefont {Tan}, \citenamefont {Han},\ and\ \citenamefont {Lim}}]{hou10}%
  \BibitemOpen
  \bibfield  {author} {\bibinfo {author} {\bibfnamefont {H.~W.}\ \bibnamefont
  {Hou}}, \bibinfo {author} {\bibfnamefont {A.~A.~S.}\ \bibnamefont {Bhagat}},
  \bibinfo {author} {\bibfnamefont {A.~G.~L.}\ \bibnamefont {Chong}}, \bibinfo
  {author} {\bibfnamefont {P.}~\bibnamefont {Mao}}, \bibinfo {author}
  {\bibfnamefont {K.~S.~W.}\ \bibnamefont {Tan}}, \bibinfo {author}
  {\bibfnamefont {J.}~\bibnamefont {Han}}, \ and\ \bibinfo {author}
  {\bibfnamefont {C.~T.}\ \bibnamefont {Lim}},\ }\href@noop {} {\bibfield
  {journal} {\bibinfo  {journal} {Lab Chip}\ }\textbf {\bibinfo {volume}
  {10}},\ \bibinfo {pages} {2605} (\bibinfo {year} {2010})}\BibitemShut
  {NoStop}%
\bibitem [{\citenamefont {Loomis}\ \emph {et~al.}(2010)\citenamefont {Loomis},
  \citenamefont {McNeeley},\ and\ \citenamefont {Bellamkonda}}]{loomis10}%
  \BibitemOpen
  \bibfield  {author} {\bibinfo {author} {\bibfnamefont {K.}~\bibnamefont
  {Loomis}}, \bibinfo {author} {\bibfnamefont {K.}~\bibnamefont {McNeeley}}, \
  and\ \bibinfo {author} {\bibfnamefont {R.~V.}\ \bibnamefont {Bellamkonda}},\
  }\href@noop {} {\bibfield  {journal} {\bibinfo  {journal} {Soft Matter}\
  }\textbf {\bibinfo {volume} {7}},\ \bibinfo {pages} {839} (\bibinfo {year}
  {2010})}\BibitemShut {NoStop}%
\bibitem [{\citenamefont {Cabral}\ \emph {et~al.}(2011)\citenamefont {Cabral},
  \citenamefont {Matsumoto}, \citenamefont {Mizuno}, \citenamefont {Chen},
  \citenamefont {Murakami}, \citenamefont {Kimura}, \citenamefont {Terada},
  \citenamefont {Kano}, \citenamefont {Miyazono}, \citenamefont {Uesaka} \emph
  {et~al.}}]{cabral11}%
  \BibitemOpen
  \bibfield  {author} {\bibinfo {author} {\bibfnamefont {H.}~\bibnamefont
  {Cabral}}, \bibinfo {author} {\bibfnamefont {Y.}~\bibnamefont {Matsumoto}},
  \bibinfo {author} {\bibfnamefont {K.}~\bibnamefont {Mizuno}}, \bibinfo
  {author} {\bibfnamefont {Q.}~\bibnamefont {Chen}}, \bibinfo {author}
  {\bibfnamefont {M.}~\bibnamefont {Murakami}}, \bibinfo {author}
  {\bibfnamefont {M.}~\bibnamefont {Kimura}}, \bibinfo {author} {\bibfnamefont
  {Y.}~\bibnamefont {Terada}}, \bibinfo {author} {\bibfnamefont {M.~R.}\
  \bibnamefont {Kano}}, \bibinfo {author} {\bibfnamefont {K.}~\bibnamefont
  {Miyazono}}, \bibinfo {author} {\bibfnamefont {M.}~\bibnamefont {Uesaka}},
  \emph {et~al.},\ }\href@noop {} {\bibfield  {journal} {\bibinfo  {journal}
  {Nature nanotechnology}\ }\textbf {\bibinfo {volume} {6}},\ \bibinfo {pages}
  {815} (\bibinfo {year} {2011})}\BibitemShut {NoStop}%
\bibitem [{\citenamefont {Shevkoplyas}\ \emph {et~al.}(2005)\citenamefont
  {Shevkoplyas}, \citenamefont {Yoshida}, \citenamefont {Munn},\ and\
  \citenamefont {Bitensky}}]{munn05}%
  \BibitemOpen
  \bibfield  {author} {\bibinfo {author} {\bibfnamefont {S.~S.}\ \bibnamefont
  {Shevkoplyas}}, \bibinfo {author} {\bibfnamefont {T.}~\bibnamefont
  {Yoshida}}, \bibinfo {author} {\bibfnamefont {L.~L.}\ \bibnamefont {Munn}}, \
  and\ \bibinfo {author} {\bibfnamefont {M.~W.}\ \bibnamefont {Bitensky}},\
  }\href@noop {} {\bibfield  {journal} {\bibinfo  {journal} {Anal. Chem.}\
  }\textbf {\bibinfo {volume} {77}},\ \bibinfo {pages} {933} (\bibinfo {year}
  {2005})}\BibitemShut {NoStop}%
\bibitem [{\citenamefont {Kumar}\ and\ \citenamefont
  {Graham}(2011)}]{kumar_seg}%
  \BibitemOpen
  \bibfield  {author} {\bibinfo {author} {\bibfnamefont {A.}~\bibnamefont
  {Kumar}}\ and\ \bibinfo {author} {\bibfnamefont {M.~D.}\ \bibnamefont
  {Graham}},\ }\href@noop {} {\bibfield  {journal} {\bibinfo  {journal} {Phys.
  Rev. E}\ }\textbf {\bibinfo {volume} {84}},\ \bibinfo {pages} {066316}
  (\bibinfo {year} {2011})}\BibitemShut {NoStop}%
\bibitem [{\citenamefont {Zurita-Gotor}\ \emph {et~al.}(2012)\citenamefont
  {Zurita-Gotor}, \citenamefont {Blawzdziewicz},\ and\ \citenamefont
  {Wajnryb}}]{zurita12}%
  \BibitemOpen
  \bibfield  {author} {\bibinfo {author} {\bibfnamefont {M.}~\bibnamefont
  {Zurita-Gotor}}, \bibinfo {author} {\bibfnamefont {J.}~\bibnamefont
  {Blawzdziewicz}}, \ and\ \bibinfo {author} {\bibfnamefont {E.}~\bibnamefont
  {Wajnryb}},\ }\href@noop {} {\bibfield  {journal} {\bibinfo  {journal} {Phys.
  Rev. Lett.}\ }\textbf {\bibinfo {volume} {108}},\ \bibinfo {pages} {68301}
  (\bibinfo {year} {2012})}\BibitemShut {NoStop}%
\bibitem [{\citenamefont {Da~Cunha}\ and\ \citenamefont
  {Hinch}(1996)}]{cunha96}%
  \BibitemOpen
  \bibfield  {author} {\bibinfo {author} {\bibfnamefont {F.~R.}\ \bibnamefont
  {Da~Cunha}}\ and\ \bibinfo {author} {\bibfnamefont {E.~J.}\ \bibnamefont
  {Hinch}},\ }\href@noop {} {\bibfield  {journal} {\bibinfo  {journal} {J.
  Fluid Mech.}\ }\textbf {\bibinfo {volume} {309}},\ \bibinfo {pages} {211}
  (\bibinfo {year} {1996})}\BibitemShut {NoStop}%
\bibitem [{\citenamefont {Smart}\ and\ \citenamefont
  {Leighton}(1991)}]{leighton91}%
  \BibitemOpen
  \bibfield  {author} {\bibinfo {author} {\bibfnamefont {J.~R.}\ \bibnamefont
  {Smart}}\ and\ \bibinfo {author} {\bibfnamefont {D.~T.}\ \bibnamefont
  {Leighton}},\ }\href@noop {} {\bibfield  {journal} {\bibinfo  {journal}
  {Phys. Fluids A}\ }\textbf {\bibinfo {volume} {3}},\ \bibinfo {pages} {21}
  (\bibinfo {year} {1991})}\BibitemShut {NoStop}%
\bibitem [{\citenamefont {Ma}\ and\ \citenamefont {Graham}(2005)}]{graham05}%
  \BibitemOpen
  \bibfield  {author} {\bibinfo {author} {\bibfnamefont {H.}~\bibnamefont
  {Ma}}\ and\ \bibinfo {author} {\bibfnamefont {M.~D.}\ \bibnamefont
  {Graham}},\ }\href@noop {} {\bibfield  {journal} {\bibinfo  {journal} {Phys.
  Fluids}\ }\textbf {\bibinfo {volume} {17}},\ \bibinfo {pages} {083103}
  (\bibinfo {year} {2005})}\BibitemShut {NoStop}%
\bibitem [{\citenamefont {Bird}(1994)}]{bird94}%
  \BibitemOpen
  \bibfield  {author} {\bibinfo {author} {\bibfnamefont {G.~A.}\ \bibnamefont
  {Bird}},\ }\href@noop {} {\emph {\bibinfo {title} {Molecular gas dynamics and
  the direct simulation of gas flows}}}\ (\bibinfo  {publisher} {Oxford
  University Press},\ \bibinfo {year} {1994})\BibitemShut {NoStop}%
\bibitem [{\citenamefont {Ivanov}\ and\ \citenamefont
  {Gimelshein}(1998)}]{ivanov98}%
  \BibitemOpen
  \bibfield  {author} {\bibinfo {author} {\bibfnamefont {M.~S.}\ \bibnamefont
  {Ivanov}}\ and\ \bibinfo {author} {\bibfnamefont {S.~F.}\ \bibnamefont
  {Gimelshein}},\ }\href@noop {} {\bibfield  {journal} {\bibinfo  {journal}
  {Ann. Rev. Fluid Mech.}\ }\textbf {\bibinfo {volume} {30}},\ \bibinfo {pages}
  {469} (\bibinfo {year} {1998})}\BibitemShut {NoStop}%
\bibitem [{\citenamefont {Bird}(1978)}]{bird78}%
  \BibitemOpen
  \bibfield  {author} {\bibinfo {author} {\bibfnamefont {G.~A.}\ \bibnamefont
  {Bird}},\ }\href@noop {} {\bibfield  {journal} {\bibinfo  {journal} {Ann.
  Rev. Fluid Mech.}\ }\textbf {\bibinfo {volume} {10}},\ \bibinfo {pages} {11}
  (\bibinfo {year} {1978})}\BibitemShut {NoStop}%
\bibitem [{\citenamefont {Drazer}\ \emph {et~al.}(2002)\citenamefont {Drazer},
  \citenamefont {Koplik}, \citenamefont {Khusid},\ and\ \citenamefont
  {Acrivos}}]{Drazer:2002gn}%
  \BibitemOpen
  \bibfield  {author} {\bibinfo {author} {\bibfnamefont {G.}~\bibnamefont
  {Drazer}}, \bibinfo {author} {\bibfnamefont {J.}~\bibnamefont {Koplik}},
  \bibinfo {author} {\bibfnamefont {B.}~\bibnamefont {Khusid}}, \ and\ \bibinfo
  {author} {\bibfnamefont {A.}~\bibnamefont {Acrivos}},\ }\href@noop {}
  {\bibfield  {journal} {\bibinfo  {journal} {J. Fluid Mech.}\ }\textbf
  {\bibinfo {volume} {460}},\ \bibinfo {pages} {307} (\bibinfo {year}
  {2002})}\BibitemShut {NoStop}%
\bibitem [{\citenamefont {Koura}(1986)}]{koura86}%
  \BibitemOpen
  \bibfield  {author} {\bibinfo {author} {\bibfnamefont {K.}~\bibnamefont
  {Koura}},\ }\href@noop {} {\bibfield  {journal} {\bibinfo  {journal} {Phys.
  Fluids}\ }\textbf {\bibinfo {volume} {29}},\ \bibinfo {pages} {3509}
  (\bibinfo {year} {1986})}\BibitemShut {NoStop}%
\bibitem [{Note1()}]{Note1}%
  \BibitemOpen
  \bibinfo {note} {$\nu = 0.5 \DOTSI \intop \ilimits@ _{0}^{L} \DOTSI \intop
  \ilimits@ _{0}^{H} n(y) (\DOTSI \intop \ilimits@ _{y-\delta _{\protect
  \textnormal {cut}}}^{y+\delta _{\protect \textnormal {cut}}} n(y+\delta )
  \protect \tmspace +\thinmuskip {.1667em} \protect \mathaccentV
  {dot}05F{\gamma } |\delta | \protect \tmspace +\thinmuskip {.1667em} d \delta
  ) \protect \tmspace +\thinmuskip {.1667em} dy \protect \tmspace +\thinmuskip
  {.1667em} dx$}\BibitemShut {NoStop}%
\bibitem [{\citenamefont {Kumar}\ and\ \citenamefont
  {Graham}(2012{\natexlab{b}})}]{kumar_jcp}%
  \BibitemOpen
  \bibfield  {author} {\bibinfo {author} {\bibfnamefont {A.}~\bibnamefont
  {Kumar}}\ and\ \bibinfo {author} {\bibfnamefont {M.~D.}\ \bibnamefont
  {Graham}},\ }\href@noop {} {\bibfield  {journal} {\bibinfo  {journal} {J.
  Comput. Phys.}\ }\textbf {\bibinfo {volume} {231}},\ \bibinfo {pages} {6682}
  (\bibinfo {year} {2012})}\BibitemShut {NoStop}%
\bibitem [{sup()}]{supmat}%
  \BibitemOpen
  \href@noop {} {}\bibinfo {note} {See the supplemental material for details on
  the calculation of pair collision and migration velocity results, comparison
  of the HMC and the BI results, and additional results on the effect of volume fraction on
  margination}\BibitemShut
  {NoStop}%
\end{thebibliography}
\end{document}